# Modeling the LAGO's detectors response to secondary particles at ground level from the Antarctic to Mexico

R. Calderón-Ardila[2,3], A. Jaimes-Motta[1,*1], C. Sarmiento-Cano[2,3,†] M. Suárez-Durán[1,3,4], A. Vásquez-Ramírez[1], for the LAGO Collaboration[5]

[1]*Escuela de Física, Universidad Industrial de Santander, Bucaramanga, Colombia*
[2]*Universidad Nacional de San Martín (UNSAM), Buenos Aires, Argentina*
[3]*Instituto de Tecnologías en Detección y Astropartículas (ITeDA), Buenos Aires, Argentina*
[4]*Departamento de Física y Geología, Universidad de Pamplona, Pamplona, Norte de Santander, Colombia*
[5]`http://lagoproject.net/`, *see the full list of members and institutions at* `http://lagoproject.net/collab.html`
*E-mail:* `christian.sarmiento@iteda.cnea.gov.ar`

The Latin American Giant Observatory (LAGO) is an observatory spanning over Latin America in a wide range of latitudes with different rigidity cut offs and atmospheric depths. The motivation of the Observatory is to study, using Water Cherenkov Detectors (WCD), the atmospheric radiation and the space weather through the measurement of secondary particles produced during the interaction of astroparticles with the atmosphere. Here, we present the methodology for estimating the number of Cherenkov photons detected by the photo-multiplier tube in a WCD in any geographical position. To do this, the secondary particle flux of cosmic rays is calculated and a geomagnetic field correction is applied. The CORSIKA and MAGCOS programs are used. In addition, the outcome of the simulation is used as an input to the Geant4 program in order to mimic the response of the detector. As a result, the distribution of photo-electrons is obtained for the total particle flux, where its behavior resembles the characteristic signals of this kind of detector and could be used to calibrate the system. This methodology was applied for five LAGO detector sites, located at different latitudes and altitudes above sea level, between Chile and Mexico.

*36th International Cosmic Ray Conference -ICRC2019-*
*July 24th - August 1st, 2019*
*Madison, WI, U.S.A.*

*Speaker.
†Corresponding author

## 1. Introduction

The detection of secondary particles at the ground level is one of the techniques used in astroparticle physics. This technique can be used to study transient events such as gamma-ray burst or Forbush decreases [1], or in applications such as muon radiography [2], known also as muography. A detailed knowledge of how the secondaries are produced in the atmosphere and how they reach a geographical position is needed for this investigations.

Currently, computational tools as CORSIKA [3] and FLUKA [4] allow the estimation of the nominal flux of those particles that arrive to the ground (background radiation). This simulation is based on the measured spectra of Galactic Cosmic Rays (GCR). The estimation allows for the calculation of the detector response to the background radiation using computational models, such as Geant4 [5], and the prediction of the effect that GCR fluctuations may have on it. The integration of these computational tools makes it possible to design new observatories for GCR, and to apply models to understand how a fluctuation on the measured signal is correlated, or not, with a perturbation on the nominal GCR flux; for instance, gamma ray burst or solar-related activity.

Detecting background radiation and using it to study astroparticles and geophysical phenomena is one of the the main goals of the Latin American Giant Observatory (LAGO). This is an extended observatory of water Cherenkov detectors (WCDs) at continental scale. It covers a large range of geomagnetic rigidity cutoffs and atmospheric absorption depths [6], (see Figure 1). In this sense, LAGO is promoting training and research in astroparticle physics in Latin America, covering three main areas: search for the high-energy component of gamma ray bursts at high altitude sites, space weather phenomena, and background radiation at the ground level [7, 8].

We developed the full computational framework to estimate the signals expected at the LAGO detector sites. This framework, which includes the effect of the geomagnetic field on the GCR propagation [8], is composed by a set of individual tools, collectively named as ARTI[1].

In this document, Section 2 introduces the method implemented by LAGO to estimate the nominal background radiation, including the correction by the geomagnetic field, at five representative locations, referenced by latitude and altitude as follows (Latitude, Altitude): Ciudad de Guatemala, Guatemala (14.63°, 1490 m a.s.l.); Bucaramanga, Colombia (7.14°, 956 m a.s.l.); Quito, Ecuador ($-0.2°$, 2800 m a.s.l.); Chacaltaya, Bolivia ($-16.35°$, 5240 m a.s.l.) and La Serena, Chile ($-29.90°$, 28 m a.s.l.). The results of the estimated signal for a standard WCD, modeled using Geant4 code, are presented for each of these positions in Section 3. Finally, in Section 4, the final remarks and future perspectives are discussed.

## 2. Estimation of Cosmic Background Radiation at The Ground Level

The flux of secondary particles at the ground was calculated for each LAGO site, following the method developed in [8]. In this method, the GCR flux ($\Phi$) is calculated at an altitude of 112 km a.s.l. Here, $\Phi$ is considered as

$$\Phi(E_p, Z, A, \Omega) \simeq j_0(Z, A) \left(\frac{E_p}{E_0}\right)^{\alpha(E_p, Z, A)}, \qquad (2.1)$$

---
[1]http://wiki.lagoproject.net/index.php?title=ARTI

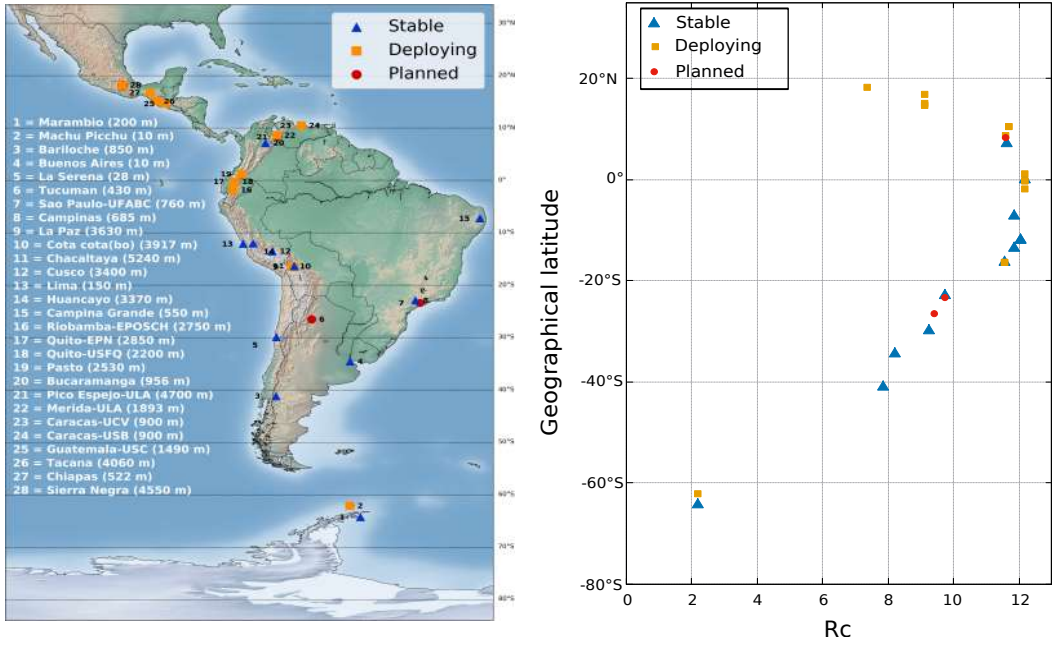

Figure 1: (left) Geographical distribution and altitudes of the Latin American Giant Observatory water Cherenkov detectors: the ones in operation are represented with blue triangles, orange squares are used for those in deployment and the planned sites are indicated in red circles. (right) Vertical rigidity cutoff at each Latin American Giant Observatory site.

where $E_p$ is the energy of the particle, $\alpha(E_p,Z,A)$ is considered constant with respect to the energy, i.e. $\alpha(E_p,Z,A) \approx \alpha(Z,A)$, from $10^{11}$ eV to $10^{15}$ eV [9], and $E_0$ has a value of $10^{12}$ eV.

In the first stage, ARTI uses CORSIKA to calculate the particles produced by the interaction of each GCR with the atmosphere [3]. Thus, we can estimate the expected flux of secondary particles at the detector level for each LAGO site.

To achieve this, the CORSIKA 76500 version was used, compiled with the following options: QGSJET-II-04,[10]; GHEISHA-2002; EGS4; curved and external atmosphere and volumetric detector. The IGRF-12 model provides the local geomagnetic field values, $B_x$ and $B_z$, required by CORSIKA to take into account the geomagnetic effects on the particles propagation in the atmosphere.

In this simulation, each secondary particle is tracked up to the lowest energy threshold that CORSIKA allows ($E_s$), according to the type of the secondary. Currently, these threshold are $E_s \geq$ 5 MeV for $\mu^{\pm}$ and hadrons (excluding $\pi^0$); and $E_s \geq$ 5 KeV for $e^{\pm}, \pi^0$ and $\gamma$. Since the atmospheric profile is a key factor for the production of secondary particles, and a parameter for CORSIKA, we have to set atmospheric MODTRAN profiles models [11] according to the geographical position of the LAGO sites: a tropical profile for Bucaramanga (BGA), Ciudad de Guatemala (GUA), Quito (UIO), La Serena (LSC) and Chacaltaya (CHA). In this way, we estimate the spectrum of secondary particles ($\Xi$).

ARTI uses input parameters such as the city code (the IATA and/or ICAO airport code[2]), time length for the flux simulation, the magnetic field, energy range, type of primary particle and the angular distribution to create a CORSIKA data file[3], needed to perform the corresponding simulations, one for each GCR.

---

[2]IATA: https://en.wikipedia.org/wiki/IATA_airport_code; ICAO: https://en.wikipedia.org/wiki/ICAO_airport_code

In this work, the flux $\Xi$ has been calculated for each of the five LAGO sites. Figure 2 shows examples of the results for the obtained spectra, for each type of secondary at CHA (5240 m a.s.l.) and LSC (28 m a.s.l.). As expected, there are less particles at low altitude due to atmospheric absorption. Namely, the flux at CHA is larger than the one at LSC for each type of secondary. Furthermore, the CORSIKA energy cuts ($E_s$) over the respective types of secondaries can also be seen in the plots.

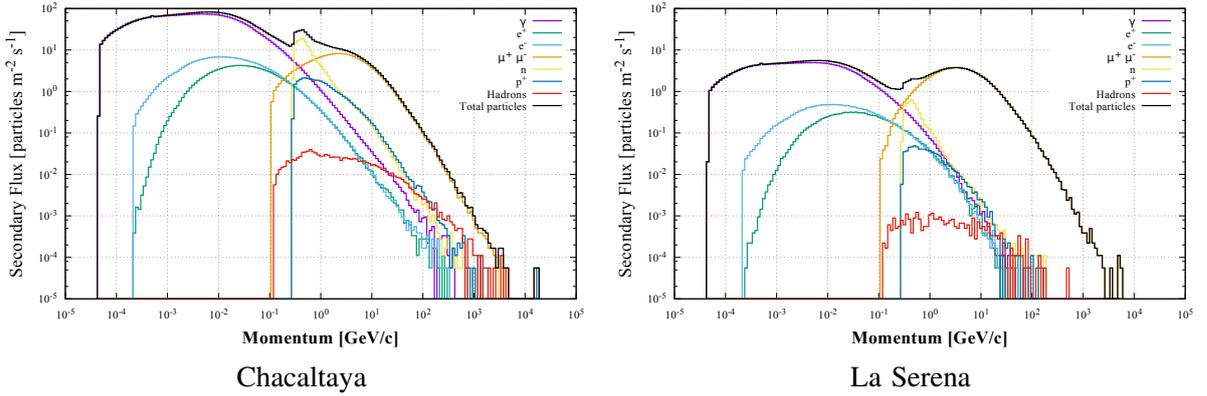

Figure 2: Spectrum of the secondary particles at two LAGO sites: (left) Chacaltaya, Bolivia (5240 m a.s.l.); (right) La Serena, Chile (28 m a.s.l.). The comparison of the two plots put in evidence a difference of one order of magnitude (at $\sim 10^{-2}$ GeV/c) in the total of secondaries (black line). At $\sim 3.5 \times 10^{-1}$ GeV/c, the neutron component (yellow line) for Chacaltaya dominates over the $\mu^{\pm}$ component (orange line), while for the La Serena estimation is the opposite. Since the LAGO detectors calibration is based on the muons, it is important to note that the prediction for the muon component is larger than for the $e^{\pm}$ (green an blue lines) ones at La Serena; meanwhile, at Chacaltaya, $e^{\pm}$ dominates with respect to $\mu^{\pm}$ due to atmospheric development of hadronic cascades.

## 2.1 Cosmic Background Radiation at ground corrected by Geomagnetic Field

Low energy GCRs ($E_p \lesssim 20$ GeV) trajectories are deflected by the Earth magnetic field (GF). The deflection is parametrized by the magnetic rigidity term ($R_m$) [12, 13, 14]. For instance, transient solar phenomena, such as Forbush decrease (FD) events, change the GF lines, the flux at low energy and, therefore, the cosmic background radiation at the ground [15, 1, 16]. The FD events have been registered by different observatories using WCDs [17, 18, 19, 20]. In this sense, the LAGO Collaboration have developed the LAGO Space Weather (LAGO-SW) program [8], to study the variations in the flux of secondary particles at ground level and their relation to the heliospheric modulation of GCRs [8]. The GF effect on the flux $\Xi$ has been included in this work for each of the five LAGO sites, following the LAGO-SW method, i.e. ignoring the secondaries produced by GCRs that do not reach the respective location. This effect is the second component of the ARTI framework.

It is important to remark that this method builds a magnetic rigidity cutoff ($R_C$) as a function of the geographical latitude, longitude, altitude above sea level, the arrival direction ($\phi$ and $\theta$) and a cumulative probability distribution function for the penumbra region, at it is explained in [8].

---

[3]ASCII file with all the parameters listed

The results for the estimated flux of cosmic background radiation at ground, including the GF correction, for the five LAGO sites are presented in Table 1 and Figure 3. Here, we can see a correlation between the flux $\Xi$ and the altitude, i.e., $\Xi$ increases with the altitude.

Table 1: Flux of cosmic background radiation at ground ($\Xi$ [m$^{-2}$ s$^{-1}$]) estimated at for five LAGO sites: Chacaltaya, Bolivia (CHA); Quito, Ecuador (UIO); Ciudad de Guatemala (GUA); Bucaramanga, Colombia (BGA); and La Serena, Chile (LSC). The flux for each secondary type is presented as follow: $e^{\pm}$ and $\gamma$ ($\Xi^{\text{EM}}$); $\mu^{\pm}$ ($\Xi^{\mu}$); neutrons ($\Xi^{\text{n}}$); and all secondaries ($\Xi^{\text{All}}$). GE$^i$ [%] represents how bigger was the GF effect over each type of secondary (estimated as the percent difference with respect to the flux $\Xi$ without GF effect), with $i$ according to the $\Xi$ notation before.

| LAGO site | Alt [m a.s.l.] | $\Xi^{\text{EM}}$ [m$^{-2}$ s$^{-1}$] | GE$^{\text{EM}}$ [%] | $\Xi^{\mu}$ [m$^{-2}$ s$^{-1}$] | GE$^{\mu}$ [%] | $\Xi^{\text{n}}$ [m$^{-2}$ s$^{-1}$] | GE$^{\text{n}}$ [%] | $\Xi^{\text{All}}$ [m$^{-2}$ s$^{-1}$] | GE$^{\text{All}}$ [%] |
|---|---|---|---|---|---|---|---|---|---|
| CHA | 5240 | 4030 | -15.4 | 231 | -12.5 | 150 | -81.3 | 4450 | -17.7 |
| UIO | 2800 | 1073 | -9.87 | 147 | -7.48 | 35.0 | -60.0 | 1263 | -11.6 |
| GUA | 1490 | 591.0 | -3.72 | 123 | -2.43 | 16.0 | -31.2 | 733.0 | -4.22 |
| BGA | 956.0 | 424.0 | -5.42 | 109 | -3.66 | 9.00 | -44.4 | 544.0 | -5.88 |
| LSC | 28.00 | 282.0 | -2.48 | 96.0 | -1.04 | 4.00 | -25.0 | 384.0 | -2.60 |

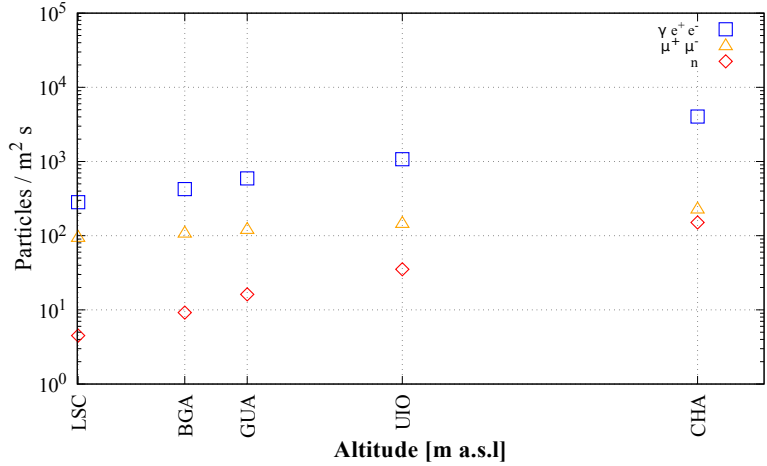

Figure 3: Flux ($\Xi$) of cosmic background radiation at ground for each LAGO site and for the different components: electromagnetic in blue squares ($\gamma$, $e^-$, $e^+$), muonic in yellow triangles ($\mu^-$, $\mu^+$) and neutrons in red diamonds (n). The flux is showed as a function of the altitude above sea level. On each component, a correlation between $\Xi$ and altitude is observed, i.e. $\Xi$, increases with the altitude as expected due to atmospheric absorption.

## 3. Estimation of the signal detected by the LAGO's WCD to the flux of cosmic background radiation

The third element of the ARTI framework is the LAGO-GD, a Geant4 [5] code that allows a detailed simulation of the interaction between the flux obtained in the previous section and the WCD. The signal detected by the LAGO detectors is estimated with this code, taking into account its geometry. The LAGO WCDs are cylindrical containers of water with an inner coating made of Tyvek® [21], and a single photo-multiplier tube (PMT, Hamamatsu R5912) at the center and top of the cylinder [22]. LAGO-GD uses the estimated flux $\Xi$ as an input parameter, distributing the number of particles on a circular area $A$, just above the WCD, during a time $t$, always conserving the flux $\Xi$.

With any cylindrical configuration (radius and height), LAGO-GD models and estimates the signal produced by Cherenkov effect as the number of photo-electrons (pe) produced in the PMT device. A pe is produced according to the quantum efficiency (QE) of the corresponding PMT (in this case, the QE from [23]). In this work, a standard LAGO WCD has been modeled with 1.05 m of radius and 0.90 m of height for all the sites. The geometry of the PMT (fully immersed in water) is taken as the photo-cathode surface with a semi-ellipsoid of semi-axis (0.101 m, 0.101 m, 0.065 m).

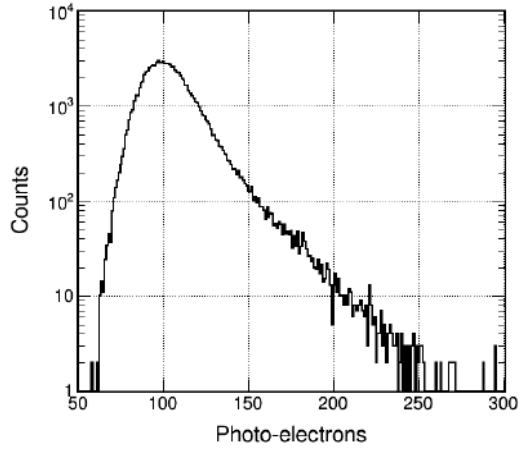

Figure 4: Distribution of the number of photo-electrons obtained for $10^5$ 3 GeV VEM using LAGO-GD. The mode of 100 pe represents the unit of calibration (1 VEM), i.e. 100 pe corresponds to $\sim$ 180 MeV of energy deposited.

The results obtained with LAGO-GD are presented according to the calibration process described in [24, 25]: converting the signal detected (in this case the number of pe) into physical units of energy deposited ($E_d$). A single high-energy muon impinging vertically at the center of the WCD, called Vertical Equivalent Muon (VEM) [24, 26], is the calibration unit defined as the average charge collected in the PMT.

In order to estimate the signal detected, the number of pe produced by VEM was calculated first. Figure 4 shows the distribution of pe obtained with LAGO-GD for $10^5$ VEMs of 3 GeV of energy, where the most probable number of pe is $\sim$ 100. This number corresponds to $E_d \sim$ 180 MeV, with a muon stopping power in water of 2 MeV/cm ; thus, our unit of calibration is 1 VEM $\sim$ 100 pe $\approx$ 180 MeV.

The charge histograms obtained for Chacaltaya and La Serena are shown in the Figure 5, where the black curve represents the total contribution of all the particles detected with the WCD. Through the simulations, it is possible to estimate the response of the WCD to different components of the Extensive Air Showers (EAS), showed in different colors. It is remarkable that the main source of the first peak is the electromagnetic component (gammas, electrons and positrons) while the second peak is dominated by the muon component. Those particles travel more distance in water, producing more Cherenkov photons than the VEM. The rate of particles detected ($\Xi^D$) and its contribution to the total energy deposited in the WCD are shown in the Table 2.

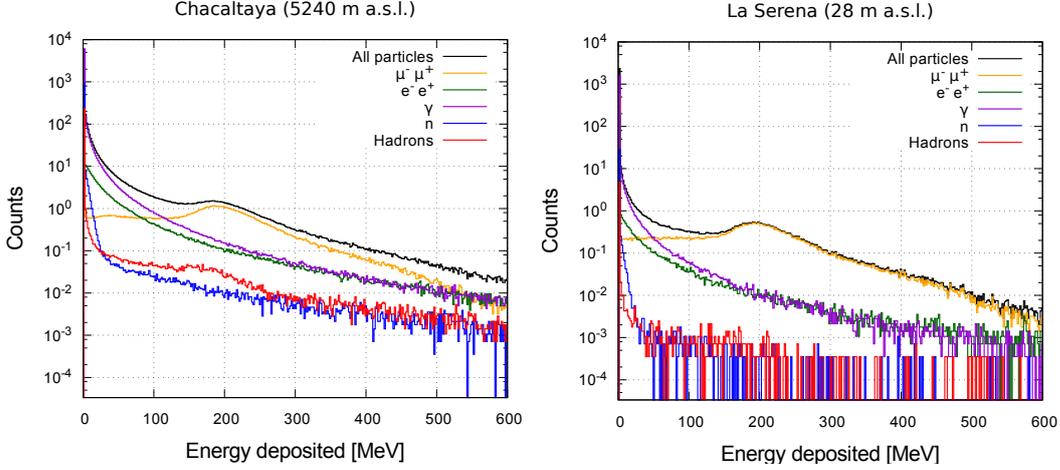

Figure 5: Charge histograms obtained for Chacaltaya (left) and La Serena (right) in an exposition time of 1 second. The black curves represent the total deposited energy and the colour ones represent the contribution of the EAS components.

Table 2: Rate of the secondary particles detected in the WCD for five LAGO sites, $\Xi^D$, and the energy deposited by the electromagnetic component, $E_d^{\text{EM}}$, muon component, $E_d^\mu$, neutrons, $E_d^n$, and all the particles, $E_d^{\text{All}}$, from $\Xi^D$.

| LAGO Site | $\Xi^D \times 10^3\,[\text{m}^{-2}\,\text{s}^{-1}]$ | $E^{\text{EM}}[\text{GeV}]$ | $E^\mu[\text{GeV}]$ | $E^n \times 10^{-1}[\text{GeV}]$ | $E^{\text{All}}[\text{GeV}]$ |
|---|---|---|---|---|---|
| CHA | 1.80 | 1.49 | 0.22 | 0.49 | 1.77 |
| UIO | 0.52 | 0.40 | 0.14 | 0.11 | 0.55 |
| GUA | 0.31 | 0.22 | 0.11 | 0.05 | 0.34 |
| BGA | 0.23 | 0.16 | 0.10 | 0.03 | 0.26 |
| LSC | 0.17 | 0.11 | 0.09 | 0.01 | 0.20 |

## 4. Conclusions

The ARTI framework, which we present here, allows us to estimate what would be the charge histogram for each site of LAGO. Therefore, those histograms can be compared with the data collected experimentally in order to calibrate the WCDs. In addition, this work agrees with the results presented in [27] where the relationship between the secondary particle flux and the height at which the detector is located was shown; see Figure 2. Finally, we were able to develop a tool that estimates the flux of secondary particles detected by a WCD in any geographic position and at any time of the year.

## 5. Acknowledgment

The LAGO Collaboration is very thankful to all the participating institutions and to the Pierre Auger Collaboration for their continuous support. In addition, some results presented in this paper were carried out using the GridUIS-2 experimental testbed, being developed under the Universidad Industrial de Santander (SC3UIS) High Performance and Scientific Computing Centre, development action with support from UIS Vicerrectoria de Investigación y Extension (VIE-UIS) and several UIS research groups as well as other funding bodies (`http://www.sc3.uis.edu.co`).

# Performance of the LAGO Water Cherenkov detector in Chiapas, Mexico

**O.G. Morales Olivares**[*,a], **Hugo de León Hidalgo**[a], **Karen Salomé Caballero Mora**[a], **Roberto Arceo Reyes**[a], **Eduardo Moreno Barbosa**[b], **Arnulfo Zepeda Domíngez**[c], **César Álvarez Ochoa**[a], **Filiberto Hueyotl Zahuantitla**[a], **Luis Rodolfo Pérez Sánchez**[a], **Elí Santos**[d] **for the LAGO Collaboration**[*].

[a]*Facultad de Ciencias en Física y Matemáticas, Universidad Autónoma de Chiapas (FCFM–UNACH)*
*Carretera Emiliano Zapata km 8, Ciudad Universitaria, Terán, Tuxtla Gutiérrez, Chiapas, C.P. 29050, México.*

[b]*Facultad de Ciencias Fisico Matemáticas, Benemerita Universidad Autónoma de Puebla (FCFM–BUAP)*
*4 Sur 104, Centro Histórico, Puebla, C.P. 72000, México.*

[c]*Centro de Investigación y de Estudios Avanzados del Instituto Politécnico Nacional (CINVESTAV)*
*Av. Instituto Politécnico Nacional 2508, Col. San Pedro Zacatenco, Gustavo A. Madero, CDMX, C.P. 07360, México.*

[d]*Mesoamerican Center for Theorethical Physics (MCTP)*
*Carretera Emiliano Zapata km 8, Ciudad Universitaria, Terán, Tuxtla Gutiérrez, Chiapas, C.P. 29050, México.*

[*] *The Latin American Giant Observatory (LAGO), http://lagoproject.net*
*See full list of members and institutions at http://lagoproject.net/collab.html*
*E-mail:* oscargmo@igeofisica.unam.mx

The Latin American Giant Observatory (LAGO) is an observatory extended along Latin America, with the capability to detect the galactic cosmic rays background and to develop studies for space weather and atmospheric radiation at ground level. It consists of a network of several Water Cherenkov Detectors (WCD) located at institutions in different countries along the American Continent (from Mexico down to the Antarctic region). One of the main goals of LAGO is to encourage and support the development of experimental basic research in Latin America, mainly with low cost equipment. The astroparticle physics group of Facultad de Ciencias en Física y Matemáticas (FCFM) at Universidad Autónoma de Chiapas (UNACH), as part of the LAGO project, is in the installation final phase of a WCD at the FCFM campus and it is planning to set up another one at the top of Tacaná volcano. In this paper, we describe the scientific purpose of the experiment, details of the detector characteristics, numerical simulations carried out to estimate its sensitivity and an electronic novelty used for the data acquisition system.

[*]Speaker.

# 1. Introduction

Nowadays astroparticle physics is one of the major fields of science due to its great contributions to the understanding of the universe, mainly those that are related to the most energetic phenomena in the cosmos. Astroparticle physics includes, beside neutrinos, gamma rays and cosmic rays. Cosmic radiation has been studied for over one hundred years since their discovery, in 1912, by Victor Hess. Water Cherenkov Detectors (WCD) have been one of the most used instruments for the study of cosmic radiation. This kind of detectors is designed to observe extensive air showers caused by the interaction of high energy cosmic rays with the Earth's atmosphere. The Latin American Giant Observatory (LAGO) is an international network of WCD, which are located in different places throughout Latin America at different altitudes, covering a wide range of geomagnetic cutoff rigidities for cosmic rays (see Figure 1).

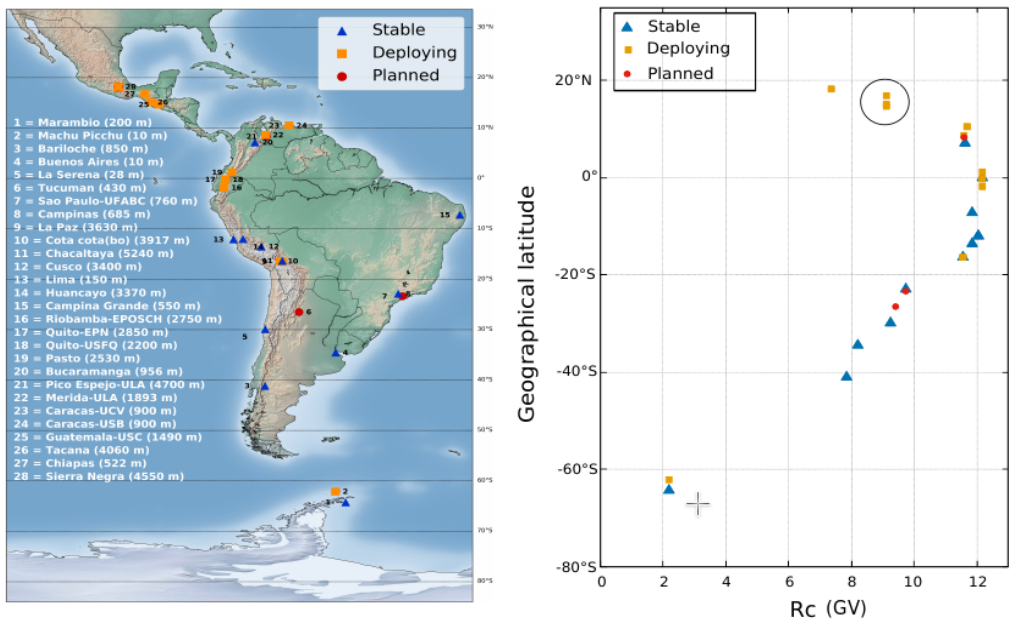

**Figure 1:** Current status of the LAGO project in each of the participant countries (left) and geomagnetic rigidity cutoff ($R_c$) for the different LAGO sites as a function of its latitude (right), the squares in the black circle correspond to the rigidity cutoff in Chiapas $\sim$ 9 GV [1].

The main scientific goals of LAGO are related to space weather, atmospheric radiation at ground level, galactic cosmic rays modulation, as well as to transient phenomena, like gamma ray bursts (GRB). It operates in an energy range that goes from 0.5 GeV to tens of TeV. As part of the LAGO project, the astroparticle physics group of Facultad de Ciencias en Física y Matemáticas (FCFM) of UNACH are making substantial efforts to install two new WCD in two sites: FCFM campus in Tuxtla Gutiérrez and the top of Tacaná volcano, both in the state of Chiapas, Mexico.

# 2. Computation of fluence at the detector's sites

To characterize the new WCD, a complete and detailed simulation is needed to get a precise estimation of the secondary flux at the detectors' positions. The simulation must take into account several important factors such as magnetic rigidity of the galactic cosmic rays (GCR), the interaction of GCRs with the atmosphere, the variations in atmospheric depth, and the response of the detector to the secondary particles. CORSIKA (COsmic Ray SImulations for KAscade) is a software that allows to perform detailed simulations of extensive air showers initiated by high energy cosmic rays [2].

Using the latest version of CORSIKA (7.6900) and the set of tools developed by the LAGO Collaboration to analyse the showers, we simulate one hour of flux at the sites: FCFM campus, at $\sim 522$ m a.s.l, and the top of Tacaná volcano, at $\sim 4060$ m a.s.l. For the simulations we need to set the geographical coordinates and the horizontal ($B_x$) and vertical ($B_z$) components of the geomagnetic field. These parameters are shown in Table 1. $B_x$ and $B_z$ were calculated with IGRF-12 model by using the NOAA–NCEI Magnetic Field Calculators. For the arrivals of the primary cosmic rays we consider a zenith angle in the range $0 < \theta < 90$ and an azimuthal angle in the range $-180 < \phi < 180$. The hadronic interaction models used at high and low energies were QGSJETII-4 and GHEISHA, respectively (see the CORSIKA manual for more details [2]). The distribution of primaries at the top of the atmosphere was calculated according to [3], taking into account the measured spectra for nuclei with atomic number within the range $1 \leq Z \leq 28$ [4]. In Figure 2, we show the fluence at the detector level for the FCFM campus and in Figure 3 for the top of the Tacaná volcano site. The results are as expected, according to the calculations for other sites of LAGO [5].

| Site | Geographical Coordinates | $B_x$ ($\mu$T) | $B_z$ ($\mu$T) |
|---|---|---|---|
| FCFM campus | 16°45′11″ N, 93°06′56″ W | 27.521 | 27.211 |
| Tacaná volcano | 15°07′48″ N, 92°06′45″ W | 27.631 | 25.598 |

**Table 1:** Geographical coordinates and horizontal and vertical components of the geomagnetic field, for the sites where the detectors will be deployed.

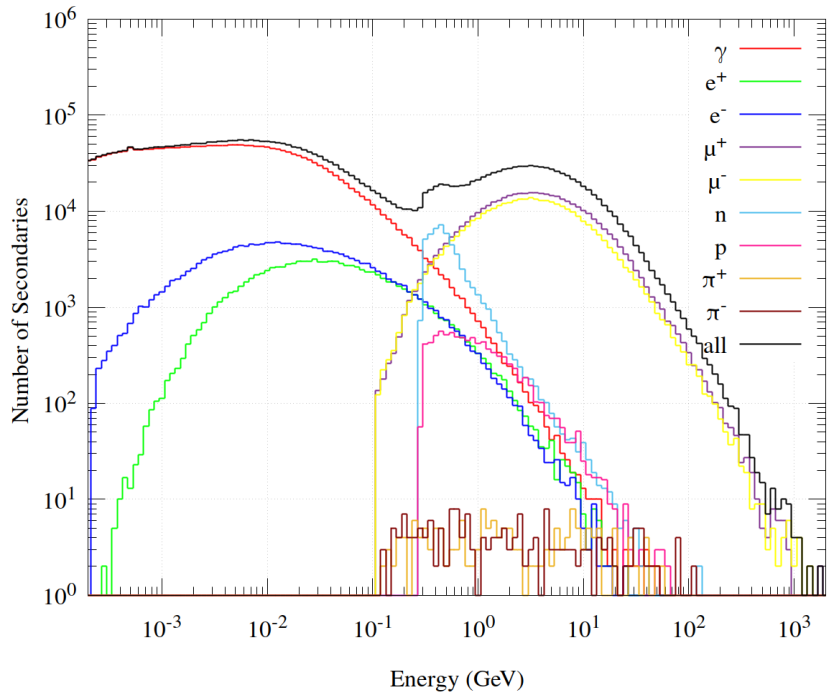

**Figure 2:** Flux of secondaries at FCFM campus at an altitude of 522 m a.s.l.

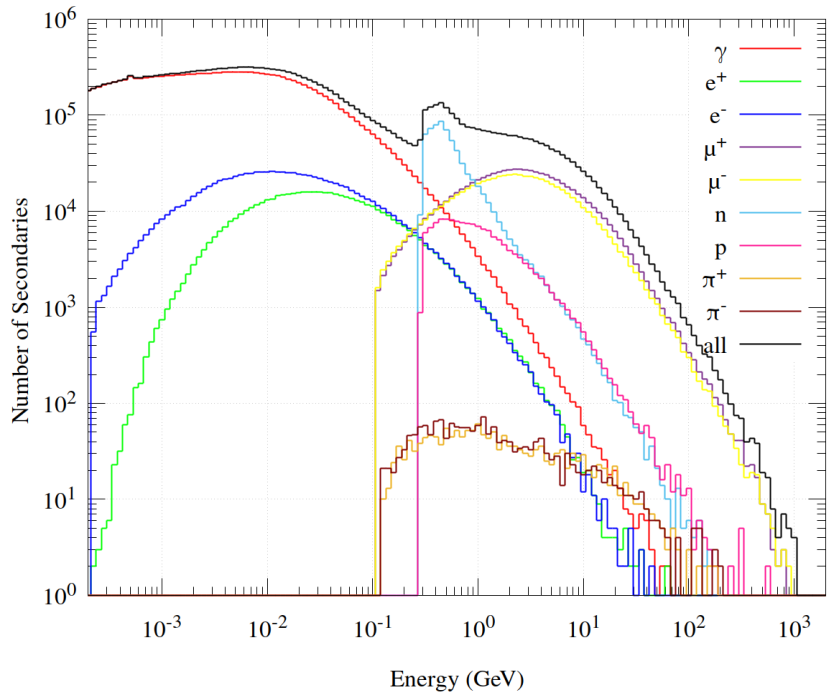

**Figure 3:** Flux of secondaries at Tacaná volcano at an altitude of 4060 m a.s.l.

## 3. Advances in the performance of the WC detector LAGO-Chiapas

### 3.1 Electronics

With respect to the electronics of the WC detector at FCFM campus, a high voltage divider circuit for polarizing the PMT has been implemented. The data aquisition system (DAQ) is achieved through a digitizer (VME1720 of CAEN) and the high voltage-polarizing card (VME1633), and to allow the PC communication with the digitizer and the high voltage card, we use a bridge car (VME1718) [6]. The DAQ includes a C-program that uses an Application Programming Interface (API), provided by the manufacturer; this program was modified according to our needs. The events are stored in an output file. Values of temperature and pressure are incoporated to this file using a BMP280 sensor. In addition, a GPS Ublox Neo 6M provides the timing labels for the events. The divider circuit was protected by a PVC-pipe case and isolated by a silicon substance. An accesory was made in order to uphold the PMT in the top of the tank. The tank is covered inside by a reflective paper and outside by a plastic material to isolate it from light.

### 3.2 The signal processing

The output file is adequately formatted so that it can be processed by the software ANNA (see http://www.iafe.uba.ar/wikilago/index.php?title=ANNA for more detailes), a set of tools developed by the LAGO Collaboration for data analisys. ANNA performs several analysis over the data, as charge and peak histograms, solar analysis, etc. ANNA had to be tailored to process our data. The main reason for this was the length of each event. It processes events of 12 bins (an aquisition window of around 300 ns, for a FPGA operating at 40 MHz). In our case 75 bins are acquired for each event (the digitizer operates at 250 MHz). Other changes were the position of the Offline trigger and the baseline given by the digitizer (2136 mV) different from the 0V used by ANNA; among others changes. Characterization of linearity, gain, single photomultiplier, etc. are being done as the last step in the process of installation of one detector in the FCFM campus, UNACH.

## 4. Conclusions

The installation of the WCD at FCFM campus is almost complete and in a near future it will be in operation to join the LAGO network. The remaining short term work regarding the simulations is to perform a Geant4 simulation of the detector to know its response to secondary particles arriving at the sites. Studies on dark current for the PMT are also ongoing.

## 5. Acknowledgements

The LAGO Collaboration is very thankful to all the participating institutions and to the Pierre Auger Collaboration for their continuous support. Furthermore, we also want to thank the high performance computing center LARCAD–UNACH (http://larcad.mx/), were the simulations presented in this work were carried out.

# Water Cherenkov detector optimization for space weather studies in the Antarctic

**L. Otiniano**[1*]**, J. Peña**[2]**, J. Vega**[1]**, V.B. Valera**[1] **and C. Castromonte**[1] **for the LAGO Collaboration**[3]**,**

[1]*Dirección de Astrofísica Comisión Nacional De Investigación y Desarrollo Aeroespacial*
*Lima, Perú*

[2]*Universidad Industrial de Santander, Bucaramanga, Colombia*

[3]*lagoproject.net/collab.html*
*E-mail:* `lotiniano@conida.gob.pe`

During the last two Peruvian summer campaigns in the Antarctica (2018 and 2019), two separate prototype water Cherenkov detectors of the LAGO project have been tested. As at high latitudes the lower geomagnetic shielding allows for the observation of cosmic rays lower than those at middle latitudes, the installation of a detector in the Antarctica is important for the development of the LAGO space weather program.

The objectives fulfilled in the second summer campaign were: a general improvement of the design of the detectors, wi-fi data transmission, autonomous operation, the elimination of light leaking and the optimization of the geometry for better separation of the electromagnetic / muonic components measured in the detector.

A comparison between the data obtained in the two Antarctic campaigns is shown, making emphasis in the measured flows and the strategy of using a second trigger level in the acquisition stage in order to perform a pre-processing of the data.

*Speaker.

## 1. Introduction

The Latinoamerican Giant Observatory (LAGO) is a network of Water Cherenkov Detectors (WCD) operated by a non-centralized collaborative network of Universities and Research Institutes in Latin America and Spain. This network is located in Latinoamérica at 7 diferent (operational) sites with altitudes, ranging from sea level (Lima, Perú) up to 5000 meters above sea level (Chacaltaya, Bolivia), and latitudes, which span most of Latin America from 18° 59' N (Sierra Negra, Mexico) to 41° 09' S (Bariloche, Argentina). The network covers an extensive range of geomagnetic shielding, and different levels of absorption and reaction at the atmosphere [1].

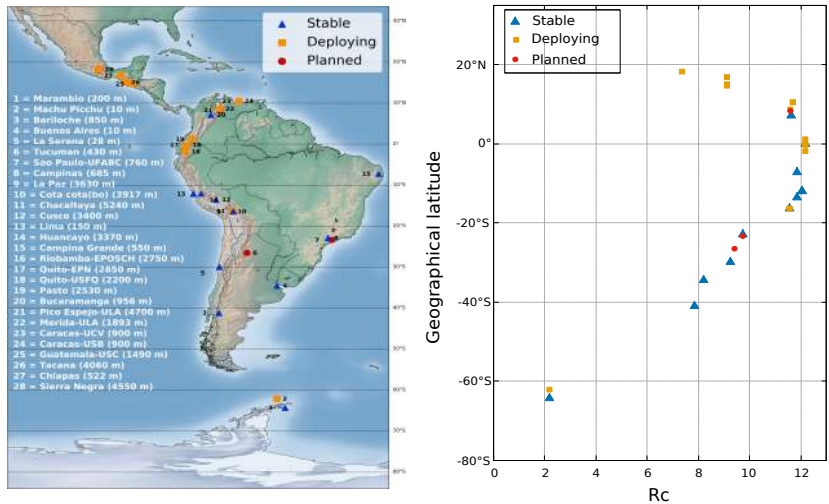

**Figure 1:** Geographical distribution and altitudes of the operational LAGO water Cherenkov detectors (bluetriangles), and those that are being deployed and will start its operation during the 2020-2021 biennium (redsquares). At the right panel, the vertical rigidity cut-off of each LAGO site is shown.

The Space Weather program of LAGO [4] is oriented to the study of the variations of the flow of secondary particles at ground level, that reflect the short and long term modulations of the galactic cosmic rays, providing information of the near space environment to Earth. Due to the geomagnetic shielding, particle detectors located at high latitudes allow the observation of cosmic rays (CRs) with lower energies than those located at middle or low latitudes. Thus, Antarctica is a privileged place to study CRs having the lowest energies that can be observed from ground level [3]. Actually LAGO project is development two sites: Argentine base Marambio [3] and Peruvian base Macchu Pichu. Here we describe the developments in detector optimization realized by the peruvian group and compare the test performed during the last two peruvian campaigns to Antartica (ANTAR XXV and ANTAR XXVI).

## 2. Antarctica Deployment

Peruvian campaign to Antarctic develops at summer season in January and February at the Macchu Picchu Scientific Station (ECAMP from spanish), it is located at Lat. 62° 05' and Long.

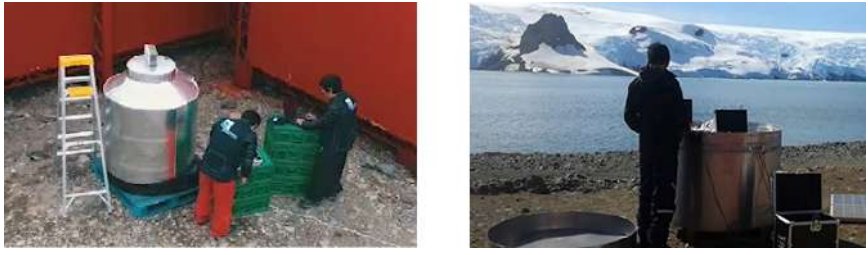

**Figure 2:** Deployment of WCD near peruvian Antartic station. Left: first detector, 2018. Rigth: second detector, 2019

58° 28' W, at 15 meters above sea level. Principal problem at the station for deplyment is the lack of stable and continuous electric power and also the lack of space that the intense scientific activity demands in the short time of realization. During the XXV peruvian campaign (2018) we test our first WCD detector outside the station.

The first WCD of the LAGO site was manufactured with commercial cylindrical plastic tank containing a volume of 1200 l ofpurified water, on top a large photomultiplier tube collects the Cherenkov radiation produced in the water by secondary CRs traversing the detector and reflected by an internal Tyvek coating. The detector is light isolated with different coatings (plastic an aluminium), see figure 2. The WCD signals are shaped and digitized by a custom made 40 MHz electronic board controlled by a Digilent Nexys2 FPGA [8].

For the second peruvian Antarctic campaign (Antar XXVI, 2019), we develop a new custom detector optimized for muon/electromagnetic separation as we show in the next session. The assembly is far more easy than previous one. The detector in made from a custom stainless steal water tank with an internal cover of $TiO_2$ based paint as substitute of Tyvek. Also the design uses and autonomous stable power supply. For protection of the operators from sudden changes in environmental conditions Wi-Fi communication with the station was necessary, see figure 2.

## 3. Optimization of the WCD

In order to calibrate the detector, we must estimate the expected charge distributions produced in the detector by Vertical Equivalent Muons (VEMs) [6]. In a previous work [5] we show a semianalitical characterization of the muon content measure in our first Antarctic detector. Based on [7] assuming a Gaussian distribution for the charge ($q$) spectrum generated by VEMs ($V(q)$) that cross a cylindrical WCD, an analytical transformation of the charge spectrum ($F(q)$) generated by atmospheric muons is performed finding that:

$$F(q) + \frac{q}{\alpha}F^{'}(q) \propto V(q) \qquad (3.1)$$

where $\alpha$ is a constant that depends on the track length distribution of muons in the detector.

Here we follow the opposite way, assuming that there is no change in $V(q)$ distribution, as first approximation, we calculate $F(q)$. In figure 3 we show this calculation, for the $V(q)$ estimation of the first detector calculated in [5].

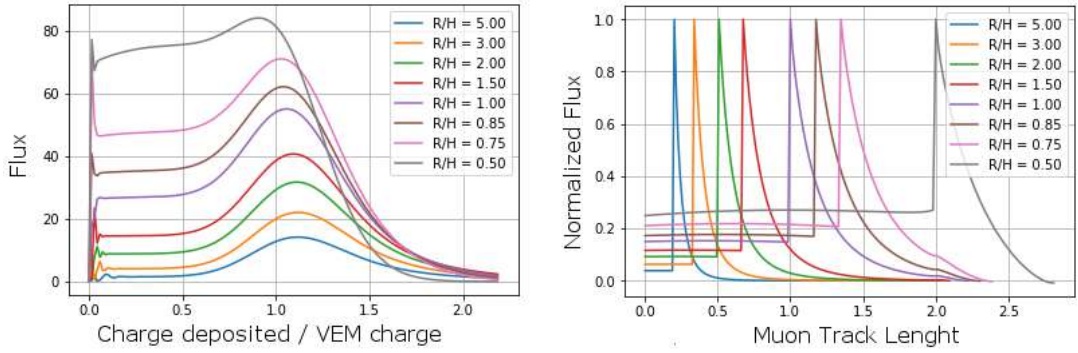

**Figure 3:** Left: expected charge distribution deposited by atmospheric muons in WCD. Right: expected track length of muon through the WCD for different of ratio R/H

The first WCD has a height H = 100 cm and radius R = 60 cm (R/H = 0.6). For this configuration, from figure 3, we expect that the muon charge distribution has a strong component of muons with a track lenght minor from vertical. To improve the muon peak we choose a ratio R/H=1 for our second WCD (H = 100 cm and radius R = 60 cm).

The signal produced by high energy particles going through the WCDs is a negative pulse with a sharp rise time of ∼10 ns and a decay time of ∼70 ns produced by the attenuation length of Cherenkov photons in the detector. The LAGO acquisition system allows to digitize the pulses in a 10 bit resolution ADC (976 $\mu$V per bin) at 40 MHz, packing the data in one hour files. The charge histogram of a WCD is obtained by time integration of the individual pulses measured in the WCD (in ADC units). As we expect that cosmic rays produce pulses with at least three bins of time we introduce a secondary trigger for the next bin (after trigger) at adquisition level with the same value of primary trigger (65 ADC), see figure 4. This reduce the data files size from 200 Mb per hour to 20 Mb. As internet data transmission is poor at ECAMP this is quite an improve for the operation.

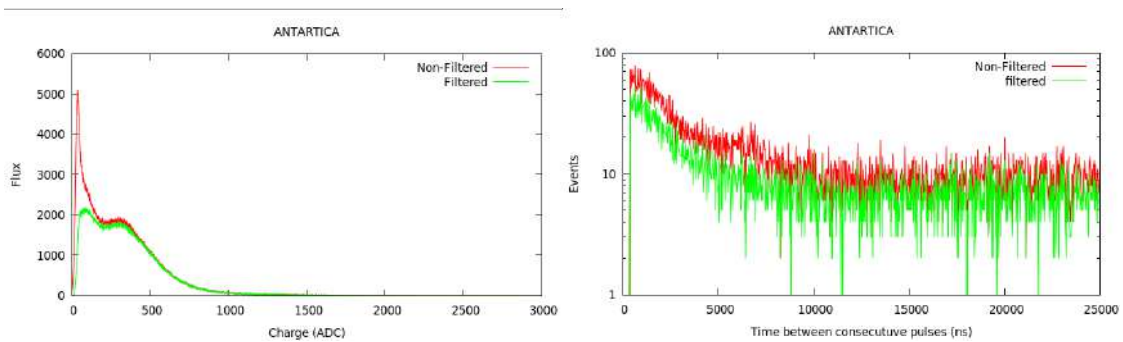

**Figure 4:** Right:Measured secondary cosmic ray charge spectrum (one hour) in the second detector with secondary trigger (green) and an with out (red)

In figure 5 we performed a muon selection (green) in the total charge spectrum (red) following the criteria used in [9]. Between two consecutive pulses, the one with the highest charge generated in the detector by secondary cosmic rays is chosen. The results from this selection are in good agreement with the estimations showed in figure 3 for both detectors.

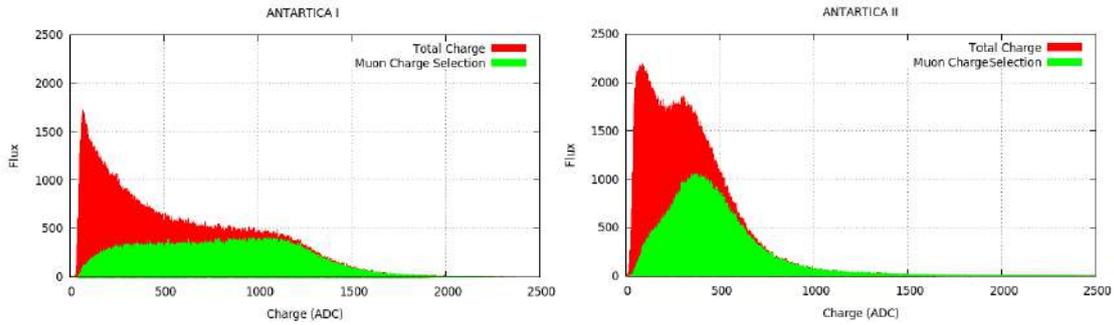

**Figure 5:** Measured secondary cosmic ray charge spectrum (one hour) and muon charge distribution estimation on Antarctic Peru's WCDs in red and muon selection applied (green)

LAGO calibration method looks for the second maximum in a charge histogram collected in one month and performs a Gaussian fit in the vicinity of this maximum, identifying this fitting function as the VEM distribution, where a central muonic band is defined. The limits in the band are given by the expected charge of muons that travel the maximum possible path in the detector (upper limit) and the transition point (local minimum) between the electromagnetic and the muonic regions of the spectrum. So it is very important that muon charge distribution to have a well resolved peak.

## 4. Conclusion and Perspectives

A method for fast evaluation of muon component in charge expectrum of WCD is under development. This method relies on a previous knowledge of the muon flux that traverse the detector and its geometry. But must been validated by simulations and independent measurements using secondary detectors ongoing. The method have been applied in the two detectors development for Antartica peruvian site of the LAGO network.

## 5. Acknowledgments

The autors gratefully acknowledges the financial support from Innovate-Perú: Fondo para la Innovación, Ciencia y Tecnología (FINCyT) trough the project PIBA-2-P-020-14 and the Ministerio de Relaciones Exteriores del Per ú through its Antarctic Affairs Direction. The LAGO Collaboration is very thankful to the Pierre Auger Collaboration for its continuous support.

# Preliminary results of the design and development of the data acquisition and processing system for the LAGO Collaboration

**L. H. Arnaldi**[*,1], **D. Cazar**[2], **M. Audelo**[3] **and I. Sidelnik**[4] **for the LAGO Collaboration**[5,†]

[1] *Centro Atómico Bariloche and Instituto Balseiro, Comisión Nacional de Energía Atómica (CNEA), S. C. de Bariloche-Argentina*

[2] *Colegio de Ciencias e Ingenierías "El Politécnico", Universidad San Francisco de Quito, Quito - Ecuador*

[3] *Facultad de Mecánica, Escuela Superior Politécnica de Chimborazo (ESPOCH), Riobamba - Ecuador*

[4] *Centro Atómico Bariloche and Instituto Balseiro and Consejo Nacional de Investigaciones Científicas y Técnicas, S.C. de Bariloche, Rio Negro - Argentina*

[5] *The Latin American Giant Observatory (LAGO), http://lagoproject.net/.*
*See full list of members and institutions at http://lagoproject.net/collab.html.*

*E-mail:* arnaldi@cab.cnea.gov.ar

We present the preliminary results obtained in the development of the new data acquisition system (DAQ) that will be used by the LAGO Collaboration. According to the requirements of the water Cherenkov detectors (WCD) used in LAGO, the new system must be capable of recording fast pulses ($\sim$ ns) from a photomultiplier (PMT), control the high voltage level applied to it, in addition to monitoring the atmospheric conditions in which the data were taken. We show some of the figures of merit indicating the performance of the new system working with a WCD. The DAQ system is based on a commercial board plus a custom made interface board. The implementation includes scalers, sub-scalers, an automatic baseline correction algorithm, pressure & temperature sensing, geolocalization, an external trigger and the capability to set and monitor the high voltage applied to the PMT. The flexibility in the design of the system allows to adapt it to different particle detector technologies, such as silicon photomultipliers (SiPMs), resistive plate chambers (RPC) and scintillators. Preliminary results prove the validity, reliability and high performance of the system.

---

[*]Speaker.
[†]for collaboration list see PoS(ICRC2019)1177

## 1. Introduction

The Latin American Giant Observatory (LAGO) is a collaboration that comes from the association of Latin American (LA) astroparticle researchers [1]. It started in 2006, and it was originally designed to survey the high-energy component of Gamma-Ray Bursts (GRBs) [2]. It is a network of ground-based water Cherenkov detectors (WCDs) located at different altitudes, between sea level and mountain Chacaltaya at more than 5200 m a.s.l., with a large range of geomagnetic rigidity cut-offs (RC). One of the primary objectives of LAGO is to install WCDs along with different sites in Latin America. At the moment, the Collaboration is composed of ten countries, forming a non-centralized and collaborative network of 25 institutions [3]. It is very important to notice that this is a collaboration to search for space phenomena and understand its physics, also working as an integrated network of science. LAGO operates as a connection of for many universities developing educational projects, helping students to deal with the construction of detectors, data acquisition systems, and data analysis, along with its physics interpretation.

It has been recently shown that WCDs can be used to study the Solar Modulation (SM) of Galactic Cosmic Rays (GCR) and other transient phenomena by measuring the variations of the flux of secondary particles at ground level [4]. Measuring the flux modulation of the GCR at different locations on Earth, which span over a wide range of rigidity cut-offs, using the same type of detectors can provide important information about the global structure of the magnetic cloud filling out the space environment surrounding the Earth. Astroparticle studies, in the context of GRBs [5, 6, 7], space weather and background radiation at ground level [8], are the main scientific objectives of the Latin American Giant Observatory.

Some of the key features of the data acquisition (DAQ) system of the LAGO detectors is the low budget needed to build it, the reliability and versatility. Unfortunately, the FPGA board used until now; the Nexys 2 [9] by Digilent is discontinued, making it very difficult and expensive to maintain WCDs already installed and implement new detectors systems as stated in the development plan of the Collaboration [10]. Given this situation, it was necessary to decide on a plan for replacing the old electronics, trying to take advantage of the state of the art in DAQ systems now available. In this context, the new DAQ concept of all-in-one arose, and we decided on the commercial board Redpitaya STEMLab [11], as it gives us the required flexibility and low cost.

This work describes the new data acquisition system used for measuring the physical quantities that are necessary to perform the analysis of low energy astroparticles at ground level. In section 2 we show the specification of the Redpitaya STEMLab hardware, that makes it suitable for our needs. Section 3 summarizes the architecture system and in 4 there is an overview of the daughterboard employed mostly for slow control. Section 5 shows results on measurements performed with the developed system, and in section 6 we give some final remarks.

## 2. The Redpitaya STEMLab board

The Redpitaya STEMLab board (originally RedPitaya board) is a low-cost ($\sim 400$ USD) analog signal generation/measurement electronics [11]. Figure 1 and Table 1 shows a schematic overview of the STEMLab Redpitaya board and some of the main specifications for the board, respectively. This commercial board is equipped with a fast dual-ADC and a fast dual-DAC, i.e., it is easy to use it as a data acquisition system and to generate output signals with an arbitrary phase difference simultaneously. These are the important specifications for its usage as a complete, versatile data acquisition system in the LAGO project. It can run a Linux distribution and can be controlled/accessed via several ways: browsers via PC or tablet (it can run a web server), usb-serial console, and the SSH network protocol. The user can easily write/modify the source code to control the FPGA and the CPU by using open-source software, i.e., it is possible to implement intellectual property (IP) modules on FPGA by using different programming languages, like VHDL or Verilog, and run C programs on the ARM CPU. Various open-source applications are also available.

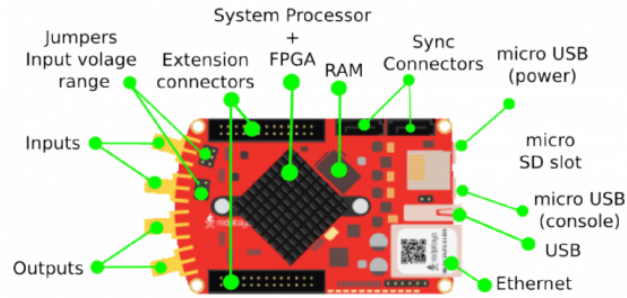

**Figure 1:** Redpitaya STEMLab board hardware overview [11].

| | |
|---|---|
| Processor | Xilinx Zynq-7010 SoC ARM dual core CPU and Artix 7 FPGA |
| System disk | MicroSD |
| RAM | 512MB |
| Access | USB console, Ethernet, WiFi dongle |
| Power | 5V x 2A max., 0.9A typ. |
| Fast DAC | Dual channel, 14-bit, 125MSPS |
| Fast ADC | Dual channel, 14-bit, 125MSPS |
| Other I/Os | Slow DAC x 4, slow ADC x 4, digital I/O x 16, daisy chain connector |

**Table 1:** Specifications of the Redpitaya STEMLab board

## 3. System architecture

The new DAQ system maintains the main characteristics of the former one [12] but incorporates new features that makes it more powerful and optimized to be cost-effective for LAGO. The principal characteristics of the two systems are summarized in Table 2. The new DAQ has a better vertical and horizontal resolution, more data bins per pulse and it runs a Linux distribution. It also has the possibility to implement a network file system (NFS) protocol for data storage and includes the monitoring of variables such as the settled high-voltage, the pulse acquisition rate, the temperature at board level and the temperature at detector level.

| Parameter | Nexys 2 based system | STEMLab based system |
|---|---|---|
| Vertical resolution | 2 mV | 122 µV |
| Horizontal resolution | 25 ns | 8 ns |
| Dynamic range | +1 V | ±1 V |
| HV monitoring | N/A | Yes |
| Rate monitoring | N/A | Yes |
| Scaler for pulse rate | Yes | N/A |
| Number of channels | 3 | 2 |
| Number of data bins | 12 (configurable) | 32 (configurable) |
| Number of data bins before trigger | 2 (configurable) | 8 (configurable) |
| NFS[1] support | N/A | Yes |
| Embedded Linux | N/A | Yes |

**Table 2:** Differences between the former LAGO DAQ system based on a Nexys 2 board, and the new one based on Redpitaya STEMLab board. (1) NFS stands for Network File System

There is a working mode where the user can set a scaler factor for the maximum frequency of pulses to be acquired. It may be useful in situations where the user needs to acquire a very low level of the signal, and the rate of pulses is huge, exceeding the maximum frequency of data acquisition in normal mode ($\sim 100$ kHz). In Figure 2, a block diagram of the DAQ system firmware (FW) is shown. FW is composed of the analog to digital converter managing block (ADC, one per channel), the baseline correction algorithm, the trigger sub-system, the external sensors managing block and all the connections to the external world. All the IP blocks comply with the AXI4-Stream standard [13], which is best suited for this application. It follows the Xilinx paradigm in constructing new designs like interconnected intellectual properties blocks.

The DAQ acquires pulses from two channels simultaneously at a sampling rate of 125 MSPS. Each channel can be connected either to one PMT or to the same PMT (anode and last dynode signals) to enhance dynamic range.

The trigger sub-system decides which pulses are stored, based on an absolute threshold settled by the user (in ADC bins) before the start of the data acquisition. In the basic configuration, a total of 32 temporal bins are stored (i.e. 256 ns at 125 MHz) for each individual pulse.

## 3.1 The baseline correction algorithm

Generally speaking, a DC offset in the input signal is undesirable because it means that the peaks of the pulses coming from the detectors are more likely to exceed the maximum input level with the consequence of a loss in information. Also, we have to take into account that any change in the level of the baseline could bias the measurements. The knowledge of the baseline level is crucial in LAGO, where we intend to measure the rate of particles at ground level with great accuracy.

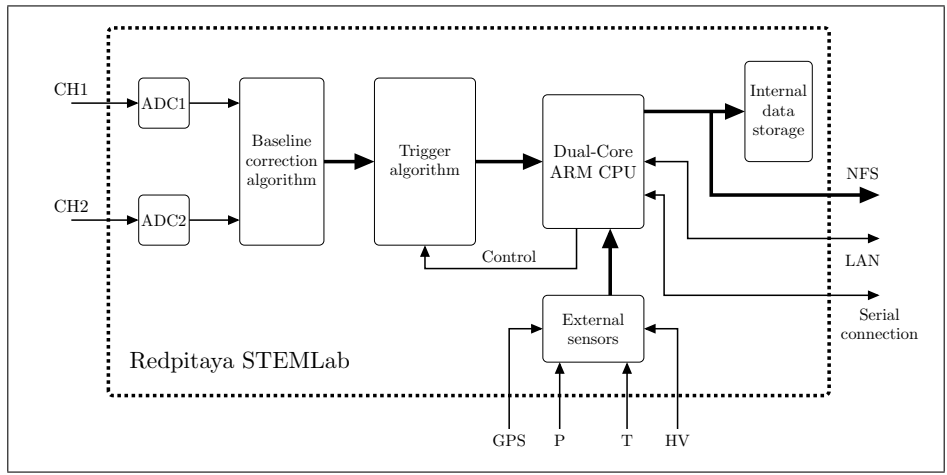

**Figure 2:** General view of the implemented system. It is shown the different IPs for signal processing and the external sensors for pressure (P), temperature (T), GPS and for high-voltage (HV) monitoring. Also, the two options for data storage, the network filesystem protocol (NFS) and the internal storage block.

DC offset should be removed before the analog-to-digital (A/D) conversion of the signal. A dedicated circuit was implemented in the former DAQ system; however, this is not possible here because STEMLab board has a fixed ADC architecture. Therefore, we explore how to remove the DC offset from the digitally sampled waveform using digital signal processing (DSP) without complicated mathematics. We investigate and apply the fractional multiplier technique [14] to create a model of a RC high-pass filter circuit and, therefore, detect the DC level of the input signal digitally and correct it.

A scheme of this implementation is shown in Figure 3, where the fractional multiplier technique is used. We have removed any multiplier in the signal path. Consequently, the baseline correction is done in a very efficient manner (in terms of resource usage and speed). The results of the application of this algorithm in the DAQ's firmware is shown in Figure 4.

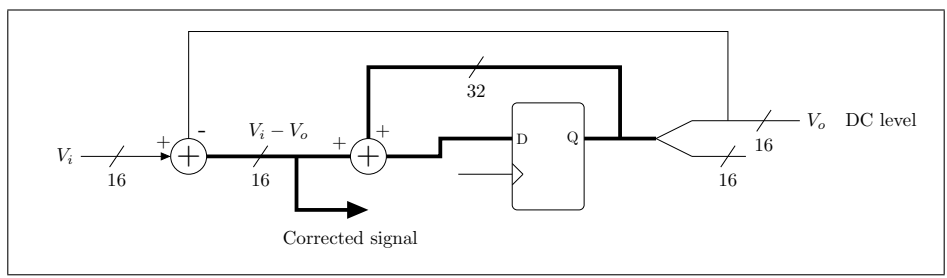

**Figure 3:** Schematic diagram showing the algorithm for the baseline correction. It is implemented as an intellectual property block in the signal's path.

## 4. The daughterboard

In order to replicate the functionality of the Nexys 2 based DAQ, a "daughterboard" has been designed and developed (Figure 5). This board provides different power supply voltages for the internal circuitry of the detectors; it powers the STEMLab board also. It replicates the STEMLab board interfaces to the external sensors and allows the system to be fed through a single external voltage source of 12 V. The board provides the $\pm 3.3$ V for the signal amplifiers in the PMT's base; the control voltage for changing the gain of the photomultiplier tube and the 12 V for the PMT's high-voltage polarization source. The inter-connection is done through the E1 and E2 connectors in the Redpitaya STEMLab board, where several standard communication protocols (SPI, I2C and UART) are available for plugging a variety of sensors to the system. Schematics, PCB and manufacturing files are available in a repository in order to allow any member of the collaboration to build her own system.

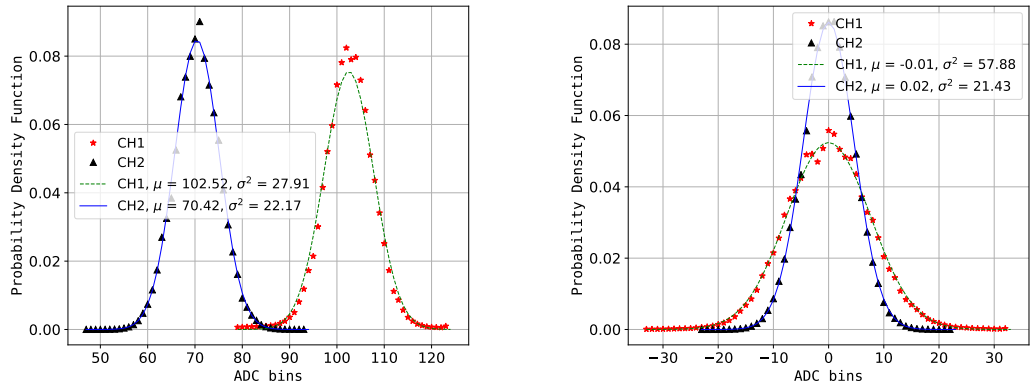

**Figure 4:** Application of the baseline correction algorithm. The plots show how the baseline correction algorithm works. It centers the baseline at the 0 ADC bin, removing any offset from it and making both channels equal. Baselines without correction (left) and corrected (right).

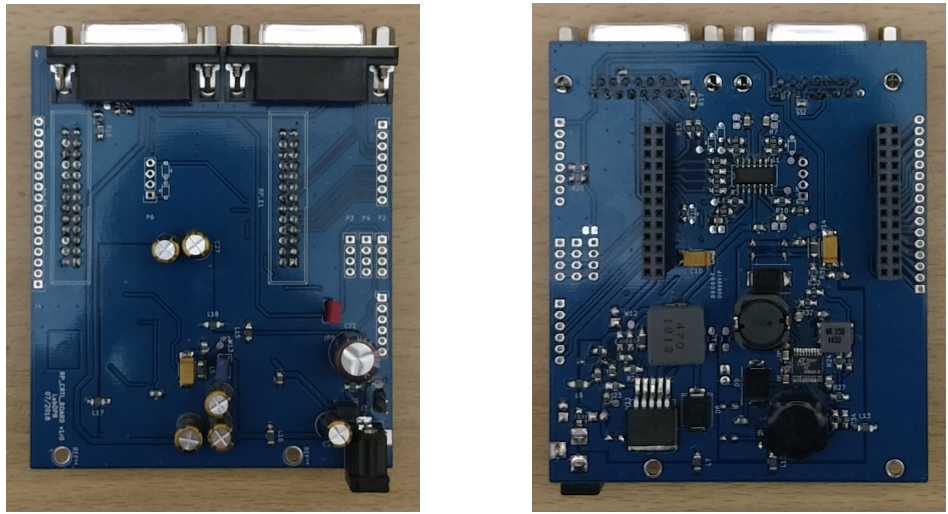

**Figure 5:** Daughter board for the Redpitaya STEMLab board.

## 5. Results of acquisition using the new DAQ

The system allows us to record different physical quantities that are of interest while measuring with the detectors. We present here some results that show the versatility of the developed system. We show temperature and pressure measurements for atmospheric monitoring and calibration of a WCD with an independent scintillator detector.

### 5.1 Coincidence measurements using independent detectors

One of the clear advantages of this system is that it can control and acquire data from two independent detectors. As a sample of this, we performed a coincidence experiment of a WCD together with a scintillator module. The WCD is a standard LAGO detector with a 9" Photonis XP1805 PMT, and the plastic scintillator module has a 1/2" Hamamatsu R1463 PMT. A diagram of the coincidence scheme can be seen in Figure 6, left. The idea is to test the capability of the developed system for handling both detectors at the same time in slow control as well as in acquisition. The coincidence allows us to calibrate the WCD using traversing muons also detected in the scintillator. An average pulse can be reconstructed from the multiple pulses recorded. This average pulse gives direct information regarding the propagation of Cherenkov photons within the detector and can be used as a tracer for the water purity and the internal wall reflectivity [12] for the WCD. Figure 6, right, shows the average pulse of the two detectors. It can be seen the difference in intensity but the similar time duration. Building the charge histogram of all recorded pulses allows one to determine the position of the vertical equivalent muon (VEM), the average energy deposited by a through-going central muon, usually appearing as a well-identified peak in the charge histogram. In this situation, we plotted the charge histograms for both detectors. This is depicted in Figure 7.

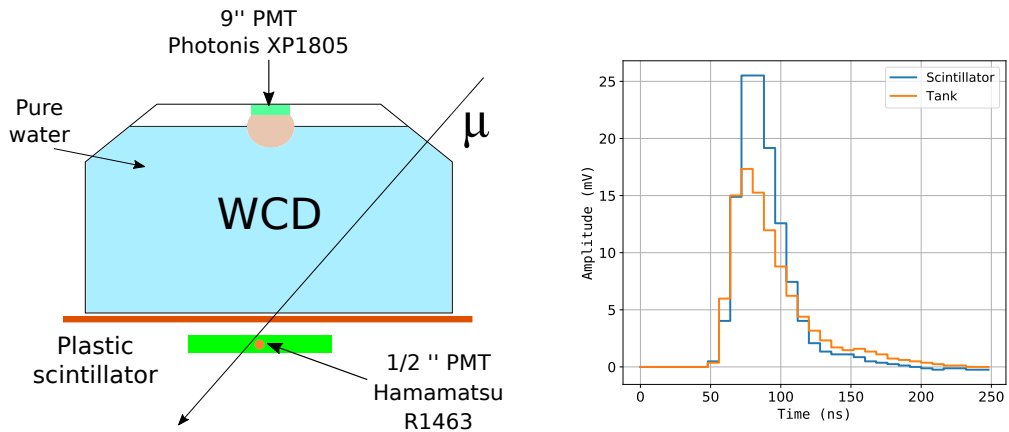

**Figure 6:** Scheme of the test measurements. There are two detectors, one is a Cherenkov tank with a photomultiplier XP1805, from Photonis and the other is composed of a plastic scintillator working with a photomultiplier R1463, from Hamamatsu (left). The mean pulse from both detectors as seen after an hour of data taking (right).

## 6. Conclusions

In this work, we have shown the upgrade of the data acquisition system used by the LAGO

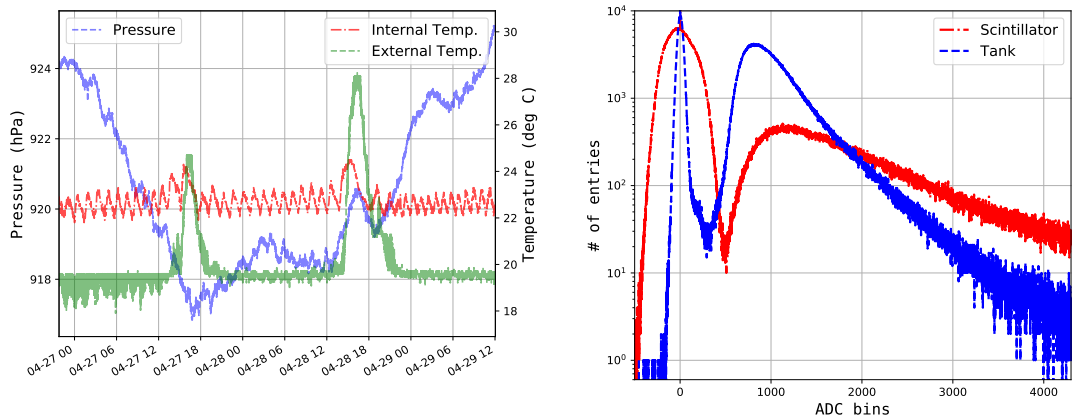

**Figure 7:** Atmospheric data sensing. The figure shows two days of data taken with the data acquisition system. The different span of the internal and external temperature can be seen (left). Charge histogram of measurements performed with both WCD and scintillator detector (right).

Collaboration. It combines the high resolution ADC found at the Redpitaya STEMLab board and a customized daughter board with the flexibility offered by an FPGA. We have shown how easy it is to adapt it to different particle detector technologies and demonstrated the feasibility and potential of this approach, opening the way for a new generation of data acquisition systems in the field of astroparticle physics. We have shown the main features of the developed system, in addition to some examples of its usage in data analysis.

## Acknowledgements

The authors would like to acknowledge the full support by CONICET and CNEA. This work has been done thanks to the following grants: PICT ANPCyT 2015-2428, PICT ANPCyT 2015-1644, PICT ANPCyT 2016-2096 and UNCuyo Proy. Cod. 06/C483.

The LAGO Collaboration is very thankful to all the participating institutions and to the Pierre Auger Collaboration for their continuous support.